\documentclass[10pt,a4paper]{article}
\usepackage{graphicx}
\usepackage{latexsym}

\begin{document}

\begin{center}
\begin{Large}\textbf{The Active Universe}\end{Large}
\begin{verbatim}
\end{verbatim}
\textit{Alexander Gl{\"u}ck$^{\ast}$, Helmuth H{\"u}ffel$^{\ast}$,
Sa{\v s}a Iliji{\'c}$^{\ast\ast}$, and Gerald Kelnhofer$^{\ast}$}\\
$^{\ast}$Faculty of Physics, University of Vienna\\
$^{\ast\ast}$Department of Physics, FER, University of Zagreb\\
{\it helmuth.hueffel@univie.ac.at, sasa.ilijic@fer.hr}
\end{center}
\begin{abstract}
Active motion is a concept in complex systems theory
and was successfully applied to various problems in nonlinear dynamics.
Explicit studies for gravitational potentials were missing so far.
We interpret the Friedmann equations with cosmological constant
as a dynamical system, which can be made `active' in a straightforward way.
These active Friedmann equations lead to a cyclic universe,
which is shown numerically.
\end{abstract}

\section{Active Motion}

To account for self-driven motion as observed in biological systems,
an additional degree of freedom, called the ``internal energy'' $e$,
was introduced in the context of complex systems theory
(\cite{schw und ebe}, \cite{schw}).
First interpretations of this formalism
were given with respect to animal movement
and complex motion of particles in general.
It describes particles who can convert their internal energy $e$
into mechanical energy and thus exhibit complex movement.
Various applications of this model were studied extensively,
e.g.\ in swarm theory \cite{swarm}.
Dynamics of a particle with position $q$ and momentum $p$
undergoing active motion as discussed in \cite{paper} are
\begin{equation}
\dot{q_i}=p_i, \ \dot{p_i}= - \frac{\partial U}{\partial q_i} \left( 1 -e \right),\label{active1}
\end{equation}
where the evolution of the internal energy $e$ is given by the equation
\begin{equation}
\dot{e}=c_1-c_2 e - c_3 e U(q).\label{active2}
\end{equation}
The $c_i$ remain constants to be fixed. The first integral of the active dynamical system reads
\begin{equation}
\frac{\dot{q}_i^2}{2}+U(q_i)-\int_0^t dt^{\prime}\ e(t^{\prime})\dot{U}(q_i(t^{\prime}))=C=\textit{const},\label{active3}
\end{equation}
which can be viewed as a generalization of the classical energy equation. In the context of active motion the particle is not simply
driven to the minimum of the potential $U(q)$, its internal energy allows it to
exhibit self-driven movement. \\ 
Depending on the structure of the potential, the systems shows various complex motion patterns, which can be studied analytically by
bifurcation theory and numerically by just simulating the evolution of $q$, $p$ and $e$ in time. So far, explicit investigations were done only
for harmonic potentials (see \cite{paper} for a detailed study of the nonlinear dynamics). One can naturally ask for active motion of particles
driven by gravitational interactions. We will now give a formulation of this problem within the framework of Friedmann equations.

\section{Active Friedmann Equations}

The Friedmann equations for the dimensionless normalized cosmological scale parameter $a(t)$, describing the spatial evolution of a flat,
homogeneous and isotropic universe with cosmological constant $\Lambda$ read
\begin{equation}
\ddot{a}=- H_0^2\left(\frac{\Omega_{r,0}}{a^3} + \frac{\Omega _{m,0}}{2a^2} - \Omega_\Lambda a  \right),\label{num}
\end{equation}
\begin{equation}
\frac{\dot{a}^2}{2}=\frac{H_0^2}{2} \left(  \frac{\Omega_{r,0}}{a^2} + \frac{\Omega_{m,0}}{a} + \Omega_\Lambda a^2 \right),\label{dup}
\end{equation}
where $\Omega_{m,0}$ and $\Omega_{r,0}$ are the density parameters of matter and radiation at present time $t=0$ and
$\Omega_\Lambda=\frac{\Lambda}{3H_0^2}$ is the density parameter of the cosmological constant. $a(0)=1$ per definition and $H_0=\dot{a}(0)$
denotes the current Hubble parameter. If we introduce the following potential
\begin{equation}
U(a)=- \frac{H_0^2}{2} \left( \frac{\Omega_{r,0}}{a^2} + \frac{\Omega_{m,0}}{a} + \Omega_\Lambda a^2 \right),
\end{equation}
then the analogy with a classical dynamical system of a single particle with coordinate $a(t)$ and energy $C=0$ is evident. According to
general relativity the constant $C$ is related to the spatial curvature of the universe which, however, is taken to be zero due to recent
observations. The density parameter of $\Lambda$ is fixed by the Friedmann equation (\ref{dup}) which at present time $t=0$ gives the relation
\begin{equation} \label{omega}
\Omega_{r,0}+\Omega_{m,0}+\Omega_\Lambda =1.
\end{equation}
We now want to propose a new phenomenological model for the development of the universe which is based on the active generalization of the
conventional Friedmann equations.
The \emph{active Friedmann equations} are given by

\begin{eqnarray}
\ddot{a} & = & - H_0^2\left(\frac{\Omega_{r,0}}{a^3} + \frac{\Omega _{m,0}}{2a^2} - \Omega_\Lambda a  \right)\left( 1- e \right), \label{num2} \\
\frac{\dot{a}^2}{2} & = & \frac{H_0^2}{2} \left( \frac{\Omega_{r,0}}{a^2} + \frac{\Omega_{m,0}}{a} + \Omega_\Lambda a^2 \right) +
\nonumber\\
& & H_0^2\int
_0^tdt^{\prime}e(t^{\prime})\dot{a}(t^{\prime})\left(\frac{\Omega_{r,0}}{a^3(t^{\prime})}+\frac{\Omega_{m,0}}{2a^2(t^{\prime})}-\Omega_\Lambda
a(t^{\prime})\right)+H_0^2C,\label{num3}\\ \dot{e} & = & c_1-c_2 e + c_3 e\frac{H_0^2}{2} \left(  \frac{\Omega_{r,0}}{a^2} +
\frac{\Omega_{m,0}}{a} + \Omega_\Lambda  a^2 \right),\label{num4}
\end{eqnarray}
where $c_1, c_2, c_3$ are arbitrary but fixed real constants. For notational convenience we have chosen $H_0^2C$ as the corresponding constant
for the first order integral in (\ref{num3}). Differing from the conventional case we allow for a non-vanishing $C$ in the active context,
which has to be determined yet. We first have to assign specific values to the density parameters of matter/radiation and the cosmological
constant. Notice that we can't set $\Omega_\Lambda=0.7$, since this results from equation (\ref{omega}), which looks different in the active
formulation, namely
\begin{equation}
\Omega_\Lambda =1-\Omega_{r,0}-\Omega_{m,0}-C,
\end{equation}
which is derived from equation (\ref{num3}) by setting $t=0$. The constant $C$ can be fixed by demanding that the current acceleration
$\ddot{a}(0)$ of the universe in the active scheme should agree with the acceleration value given by the conventional Friedmann equations (thus
accounting for the experimental result $\ddot{a}(0)>0$). For any fixed $e_0:=e(0)$ one finds
\begin{equation}
C=\frac{e_0}{1-e_0} \left( \frac{3}{2} \Omega_{m,0} + 2\Omega_{r,0} - 1 \right).
\end{equation}
Accordingly,
\begin{equation}
\Omega_\Lambda = \frac{1}{1-e_0} \left( 1-\Omega _{m,0}(1+\frac{e_0}{2})-\Omega _{r,0}(1+e_0)\right),
\end{equation}
so that $\Omega_\Lambda$ is now completely determined by the density parameters of matter/radiation and the initial condition for $e$, which is
freely chosen. Depending on this initial condition $\Omega_\Lambda$ may also be brought to vanish.

\section{Discussion}

Figure 1 shows a simulation example of the active dynamics of the normalized cosmological scale parameter $a(t)$. We observe oscillatory behavior with significantly smaller amplitudes and shorter periods in the past. The scale parameter $a(t)$ is
oscillating around a certain equilibrium point, which can be calculated analytically. The active Friedmann equations can be rewritten as
follows:
\begin{eqnarray}
\dot{a} & = & p \nonumber\\
\dot{p} & = & -\frac{dU(a)}{da}(1-e)\label{crit}\\
\dot{e} & = & c_1-c_2e-c_3eU(a).\nonumber
\end{eqnarray}
Equilibrium points are found by searching values $(\tilde{a},\tilde{p},\tilde{e})$, for which the right hand side of (\ref{crit}) vanishes. Equilibrium points with $\tilde{e}=1$ are unstable, the case $\tilde{e}\neq1$ leads to the conditions
\begin{eqnarray}
\tilde{p} & = & 0,\nonumber\\
0 & = & \Omega_\Lambda \tilde{a}^4 - \frac{\Omega_{m,0}}{2} \tilde{a} - \Omega_{r,0}, \label{crit2}\\
\tilde{e} & = & \frac{c_1}{c_2+c_3U(\tilde{a})}.\nonumber
\end{eqnarray}
For the specific choice of parameters and initial conditions made for the simulation shown in Figure 1, the numerical values of $(\tilde{a},\tilde{p},\tilde{e})$
can be calculated directly. The quartic equation for $\tilde{a}$ has only one positive, real-valued solution, namely $\tilde{a}=1.01$,
resulting in $(\tilde{a},\tilde{p},\tilde{e})=(1.01,0,4.08)$. Figure 1 shows that this value of the scale parameter marks the center point
of the oscillations. The stability analysis of the above equilibrium point $(\tilde{a},\tilde{p},\tilde{e})$ can be made by
linearizing the system (\ref{crit}) and calculating the eigenvalues of the Jacobian matrix. Our calculation shows the presence of a purely imaginary pair of eigenvalues $\lambda_{1,2}=\pm i \omega$ and we are near a Hopf bifurcation point (for bifurcation theory, see \cite{yuri}). In our case $\omega=1.19$, so that the period of oscillations is given by $T=\frac{2\pi}{\omega}=5.28$ in units of the Hubble time $H_0^{-1}$, which agrees well for large times with our numerical simulation.\\ \\

\begin{figure}
\begin{center}
\includegraphics[scale=0.6]{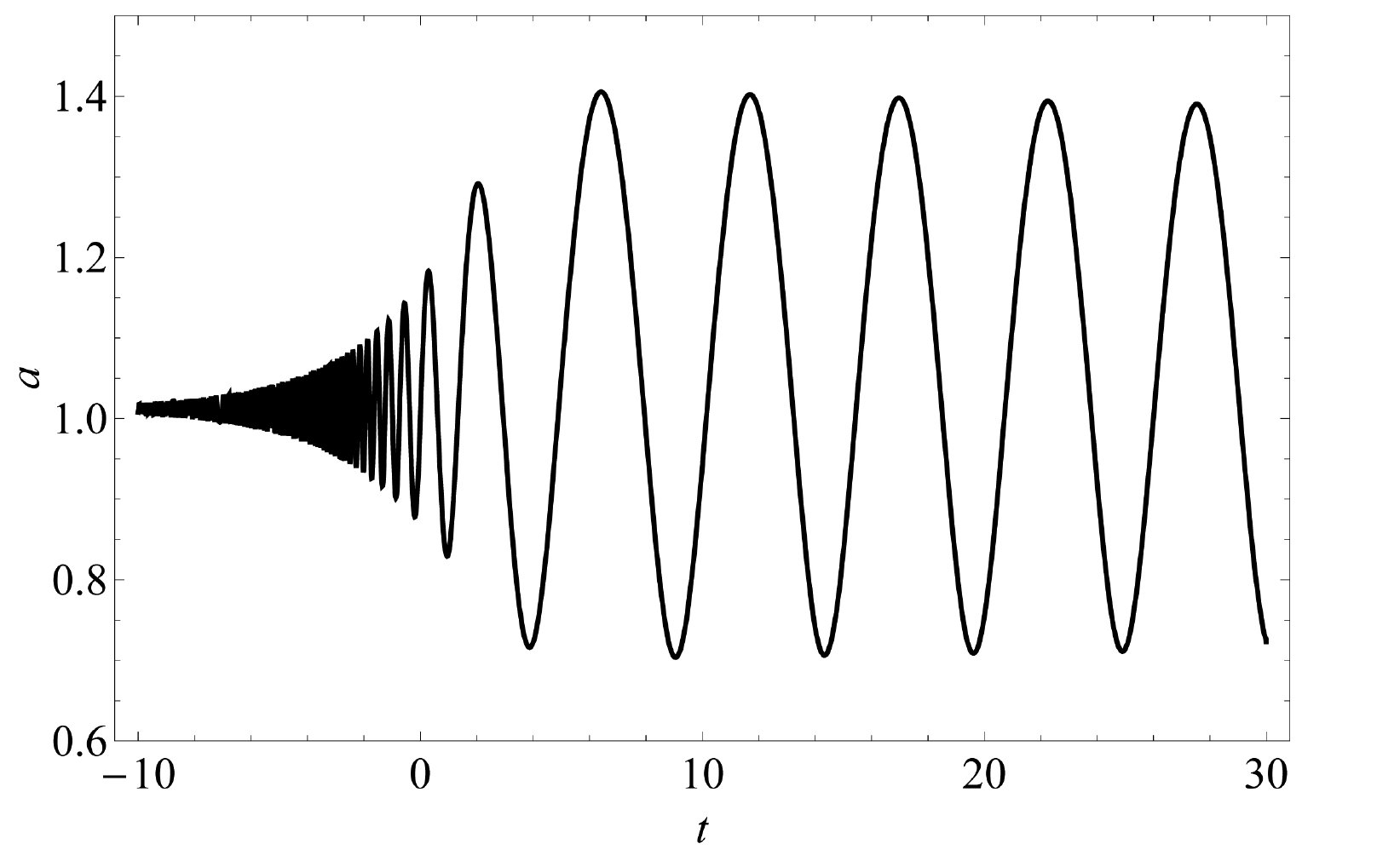}
\end{center}
\caption{Numerical result for the evolution of $a(t)$,
with $c_1=5$, $c_2=1$, $c_3=-1$, $e(0)=100$, $H_0=1$, $\Omega_{m,0}=0.29$
and $\Omega_{r,0}=0.01$.
(These are rough representative values for the density parameters,
consult \cite{cosmo} for a recent overwiev of cosmological parameters.)
This choice of parameters leads to $\Omega_\Lambda=0.15$.}
\end{figure}

Summarizing, we can say that interpreting the Friedmann equations as a nonlinear dynamical system and handling it within the framework of active motion
results in oscillatory solutions for the scale parameter. Hence we succeeded in constructing a model for a cyclic universe giving rise to a
sequence of expansions and contractions without any singularity, yet accounting for the observed spatial flatness and the current accelerated
expansion. Besides this particular example, the active Friedmann equations allow for a lot of other possible scenarios which deserve further
analysis. Whether the active universe represents a reasonable model for cosmology will depend on finding a physically meaningful interpretation
of the additional parameter $e$. Attempts can be made to identify its
effect with corrections arising e.g. from cosmological theories with extra dimensions, which will be the task of future investigations. \\ \\

\textbf{Acknowledgments}: We thank Helmuth Rumpf for valuable discussions. We are grateful for financial support within the
Agreement on Cooperation between the Universities of Vienna and Zagreb.

\end{document}